\pdfoutput=1
\documentclass[doublecol,linenumbers]{epl2} 

\usepackage{amsmath}

\title{Possible Fano resonance for high-$T_c$ multi-gap superconductivity in p-Terphenyl doped by K at the Lifshitz transition}
\shorttitle{Shape resonance in superconducting gaps in p-Terphenyl}

\author{Maria Vittoria Mazziotti~\inst{1}, Antonio Valletta~\inst{2}, Gaetano Campi~\inst{3}, Davide Innocenti~\inst{4}, Andrea Perali~\inst{5} \and Antonio Bianconi~\inst{1,3,6}}
\shortauthor{M. Vittoria Mazziotti \etal}

\institute{ 
 \inst{1} RICMASS, Rome International Center for Materials Science Superstripes, Via dei Sabelli 119A, 00185 Rome, Italy\\
 \inst{2} Institute for Microelectronics and Microsystems, IMM, Consiglio Nazionale delle Ricerche CNR, Via del Fosso del Cavaliere 100, 00133 Roma, Italy\\
\inst{3} Institute of Crystallography, IC, Consiglio Nazionale delle Ricerche CNR, via Salaria, Km 29.300, 00015 Monterotondo Roma, Italy\\
 \inst{4} Department of Chemistry, University of Liverpool, Liverpool, L69 7ZD, UK\\
 \inst{5} School of Pharmacy, Physics Unit, University of Camerino, Camerino, Italy\\
 \inst{6} National Research Nuclear University MEPhI (Moscow Engineering Physics Institute) 115409 Moscow Kashirskoe shosse 31 Russia\\}

{noncuprate superconductors }
{Theories and models of superconducting state}
{Occurrence, potential candidates}

\abstract{Recent experiments have reported the emergence of high temperature superconductivity
 with critical temperature $T_c$ between 43K and 123K in a potassium doped aromatic hydrocarbon 
 para-Terphenyl or p-Terphenyl. This achievement provides the record for the
 highest $T_c$ in an organic superconductor overcoming the previous record of $T_c$=38 K in $Cs_3C_{60}$ fulleride.
 Here we propose that the driving mechanism is the quantum resonance between superconducting gaps near a Lifshitz transition
 which belongs to the class of Fano resonances called shape resonances. 
 For the case of p-Terphenyl our numerical solutions of the multi gap equation shows that high $T_c$ 
 is driven by tuning the chemical potential by K doping and it appears only in a narrow energy range 
 near a Lifshitz transition. At the maximum critical temperature, $T_c$=123K, the condensate in the 
 appearing new small Fermi surface pocket is in the BCS-BEC crossover while the $T_c$ drops
 below 0.3 K where it is in the BEC regime. Finally we predict the experimental results which 
 can support or falsify our proposed mechanism: a) the variation of the isotope 
 coefficient as a function of the critical temperature and b) the variation of 
 the gaps and their ratios 2$\Delta$/$T_c$ as a function of $T_c$.}

\begin{document}

\maketitle

\section{Introduction}

Recently high temperature superconductivity 
with $T_c$ in the range of 43-123 K has been reported following 
different growth procedures in a potassium doped aromatic 
hydrocarbon p-Terphenyl~\cite{a,b,c}. 
The superconducting crystalline phase is expected to be $K_3C_{18 }H_{14}$. 
Superconducting pairing with a large 15 meV gap opening at about 60K 
was confirmed on a K-doped surface of a p-Terphenyl single crystal~\cite{d}. 
Para-Terphenyl, a linear molecule made of a chain of 3 benzene rings, 
and its derivatives are aromatic biological molecules present in edible mushrooms~\cite{e}.
 Pharmaceutical research is in progress for their use as
immunosuppressive, anti-inflammatory and anti-tumor
agents, moreover it has technological applications as laser dye,
sunscreen lotion and in photon detectors.
If these results will be confirmed, $K_x$ p-Terphenyl provides today the record
 for the highest critical temperature in carbon based materials and larger 
 than in many cuprates oxides and in many iron based superconductors. 
The search for macroscopic quantum coherence at high
temperature in organics and organometallics has been a
long standing search for the holy grail of room temperature superconductors. 
Superconductivity in graphite intercalation compounds
has been an active topic for several decades, but $T_c$ ranges 
only up to 11.5 K for $CaC_6$ \cite{f}. 
The highest critical temperature $T_c$=38 K in doped fullerides $A_3C_{60}$
$A=K,Rb,Cs$ has been found in $Cs_3C_{60}$ 
 with A15 structure by application of 7 Kbar hydrostatic pressure~\cite{ganin1}. 
 This system provides a complex phase of condensed matter where structural polymorphism 
 controls both magnetic and superconducting properties~\cite{ganin2} 
 and shows a fluctuating microscopic heterogeneity made of the coexistence 
 of both localized Jahn Teller active and itinerant electrons~\cite{zadik}. 
 Recently the material research for high temperature superconductors was oriented 
 toward metal-intercalated aromatic hydrocarbons by the discovery of superconductivity with $T_c$=18K
 by doping potassium into picene ($C_{22}H_{14}$)~\cite{mitsu,kubozono}. 
 Picene is a hydrocarbon molecule made of five benzene rings condensed in an armchair manner.
 Looking for high $T_c$ in alkali-metals or alkali-earth-metals doped
 polycyclic-aromatic-hydrocarbons (PAHs), the previous record 
 $T_c$=33 K was held by in $K_x$ 1,2:8,9-dibenzopentacene ($C_{30}H_{18}$)~\cite{xue}. 
The recent indications for $T_c$=123K in p-Terphenyl open a new road map for the search 
of higher critical temperatures in the large family of doped metal-organic compounds, with the hope to overcome
the highest superconducting critical temperature known so far, $T_c$=203K in pressurized $H_3S$ ~\cite{eremets,EPL}. There is today high interest on understanding both the quantum mechanism 
beyond the emergence of $T_c$=123K in p-Terphenyl and the relation between 
the nanostructure and quantum functionality in organics. This is needed to develop novel
$quantum$ $plastics$ materials, taking advantage that conducting polymers can be prepared in a large variety of polymeric 
heterostructures at atomic limit~\cite{fara,malva,botta} and possess the combination of easy processability, 
light weight, and durability. 
Using density functional theory and Eliashberg's theory of superconductivity
 in the single band approximation the superconducting critical 
 temperature has been predicted 
to be in the range $3<T_c<8$ K in $K_3$ picene~\cite{casula} and around 6.2 K 
in $K_2$$C_6$$H_6$ moreover, it was argued 
that all hydrocarbons should show $T_c$ in a similar temperature range
of $3<T_c<7$K~\cite{zhong}. A hot topic today is the search for superconductivity 
in the extreme high pressure phases of benzene~\cite{wen,katru,hof}.
Therefore unconventional pairing mechanisms were invoked for high $T_c$ in doped p-Terphenyl: 
a) the Bose-Einstein Condensation (BEC) of preformed bipolarons~\cite{a,b,c}; 
b) the Resonating Valence Bond (RVB) theory in a scenario with two coexisting doped Mott insulators~\cite{ba} 
c) the so called $s_{+/-}$ pairing mechanism mediated by a repulsive interaction in a scenario with two strongly 
correlated bands forming two similar Fermi surfaces connected by a nesting vector~\cite{fa}.

\begin{figure}
\onefigure[scale=0.8]{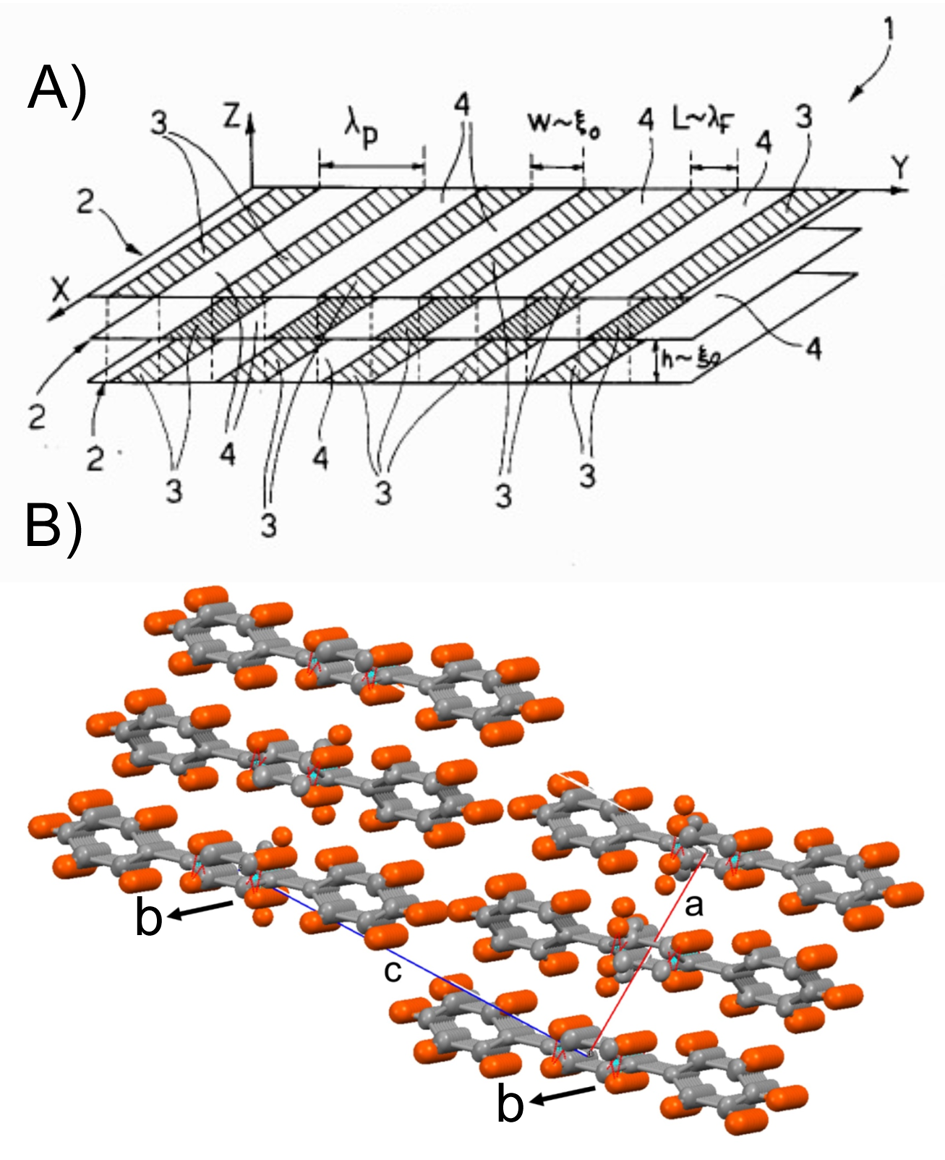}
\caption{ A) Drawing of the heterostructure at atomic limit made of nanoscale stripes which promotes shape resonances according with Ref.\cite{cuprates0}. The stripes run along the x direction of the xy planes which are separated along the z direction. B) The p-Terphenyl crystal structure P21/a (CCDC reference No. 847173) appears as a practical realization of the superlattice of stripes shown in panel (A) made of p-Terphenyl striped units of atomic thickness. In the P21/a structure the $\bf{b}$, $\bf{c}$ and $\bf{a}$ axis correspond to the x, y and z axis in the schematic drawing in panel (A). The p-Terphenyl stripes run along the $\bf{b}$ direction and are well separated by the neighbor stripe along the $\bf{c}$-axis and the $\bf{a}$-axis directions}
\label{fig.1}
\end{figure}

\begin{figure}
\onefigure[scale=0.8]{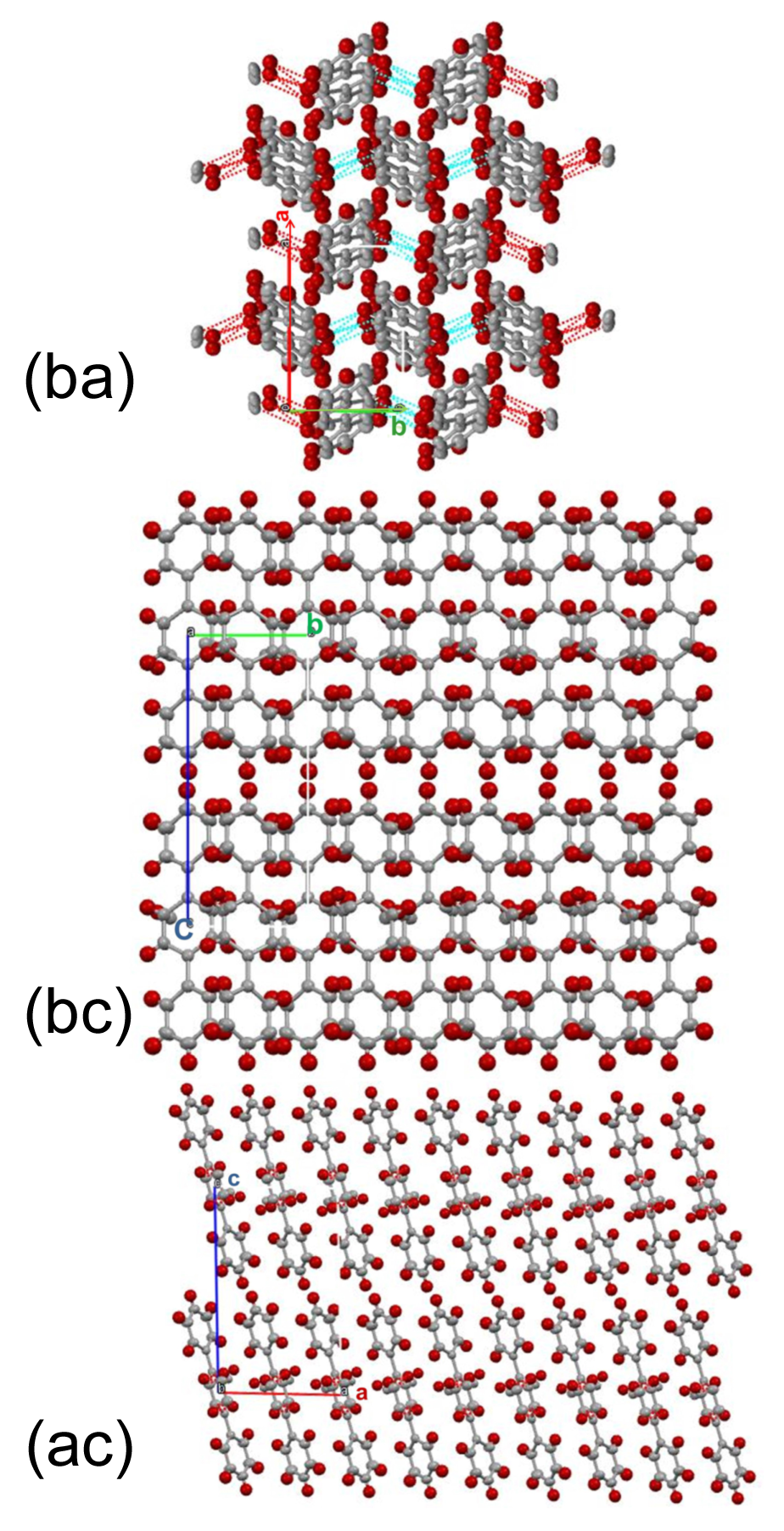}
\caption{Three different projections of the p-Terphenyl crystal structure, monoclinic 
space group P21/a, with axis $\bf{a}$=0.81 nm, $\bf{b}$=0.56 nm,
$\bf{c}$=1.36 nm and $\beta$=$92^{\circ}$.
The $\bf{bc}$ projection 
shows two parallel p-Terphenyl stripes, made of linear p-Terphenyl molecules, running in the $\bf{b}$ axis direction, 
the packing of these stripes form a superlattice 
of stripes with a period of 1.36 nm in the $\bf{c}$ direction. 
The upper $\bf{ba}$ projection shows the lateral view of the p-Terphenyl stripes.
The lower $\bf{ac}$ projection shows the packing of stripes, running in the $\bf{b}$ axis direction, 
perpendicular to the page well separated by neighbor stripe forming a superlattice of stripes 
with period of 0.81 nm in the $\bf{a}$ axis direction.}
\label{fig.2}
\end{figure}

\begin{figure}
\onefigure[scale=0.7]{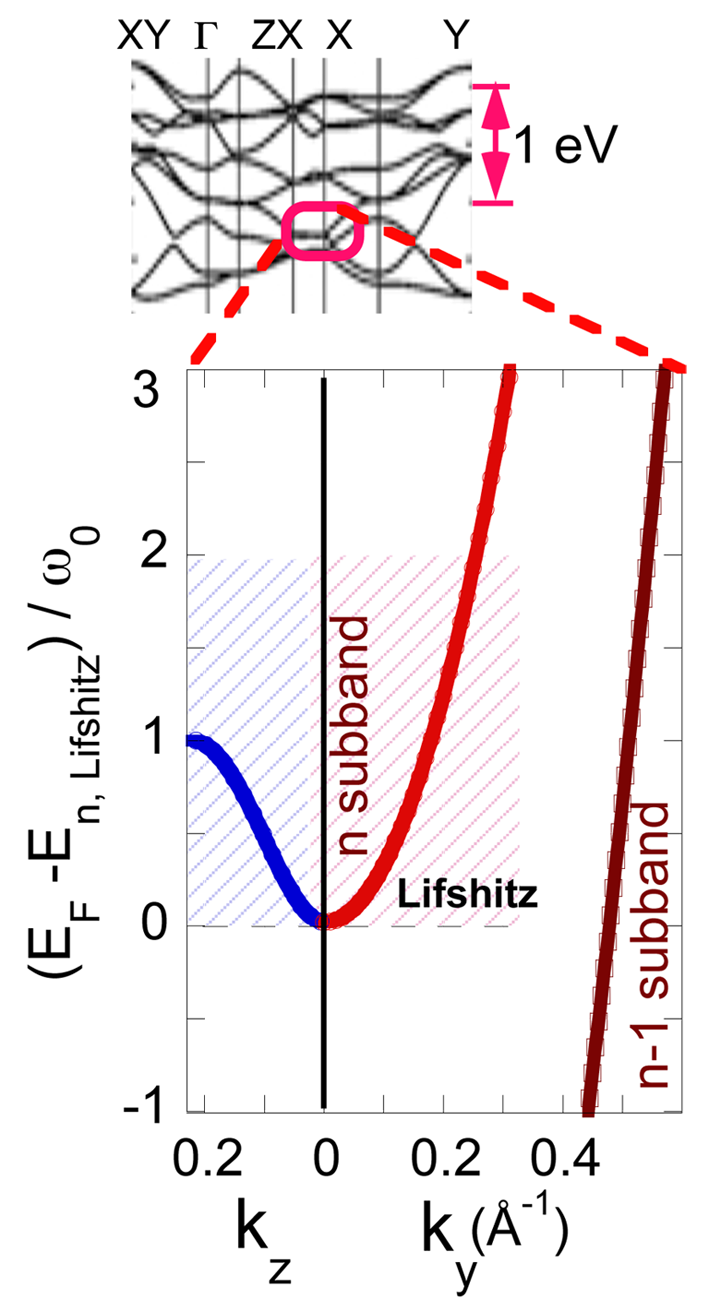}
\caption{Panel (A) shows the Brillouin zone (BZ) representation of the conduction band of the p-Terphenyl from Ref.~\cite{pusc}. The 2 eV wide conduction band is made by a bundle of many bands showing a large dispersion in the Y direction, the direction of the p-Terphenyl stripes, and flat dispersions in the ZX directions, orthogonal to the stripes direction. The red circle indicates a narrow energy range of about 600 meV around the band edge of the n-th electronic band which shows the narrow energy dispersion in the XZ direction and wide dispersion in the Y direction which coexist with other (n-1)-th dispersing bands. Panel (B) shows the dispersion of the electronic bands obtained by solution of the Schroedinger equation for a superlattice of stripes with period 1.4 nm, separated by a potential barrier, determined by the hopping energy between stripes, in the transversal direction is such that the transversal energy dispersion in the periodic potential is 145 meV the same as the assumed energy cut off for the pairing $\omega_0$=145 meV. The narrow energy range of the dome of high $T_c$ is only of the order of $4\omega_0$=580 meV, where the Fano resonance near the Lifshitz transition is in action.}
\label{fig.3}
\end{figure}

Here we propose the mechanism for $T_c$ amplification driven by Fano 
resonance~\cite{cuprates1,cuprates0} in a multigap superconductor
 between different gaps occurring where the chemical potential is tuned 
 at a electronic topological Lifshitz transition~\cite{Lifshitz,Lifshitz1}.
 In this regime the configuration interaction between the pairing scattering 
 channels in the (n-1)-th bands with high Fermi energy $E_{F(n-1)}$ in the 
 BCS approximation and the n-th pairing scattering channel in the new appearing n-th small 
 Fermi surface with a low Fermi energy $E_{Fn})$ can give a resonance 
 in the superconducting gaps, called shape resonance\cite{shape,shape2} like
 in nuclear physics and molecular physics \cite{vittorini} or Feshbach resonance 
 as in the jargon of ultracold gases~\cite{fes10}. The Fano resonance give high $T_c$ domes around Lifshitz transitions 
 in the energy range of $E_{Fn}$ of the order of the energy 
 of the pairing cut-off and the n-th condensate is in the BCS-BEC crossover~\cite{perali-bcs-bec, Guidini-bcs-bec}.

The configuration interaction between pairing channels is determined by the symmetry and interference between the
wavefunctions of electron pairs at the coexisting multiple Fermi surfaces therefore
 the nanoscale material architecture plays a central role.
In ref~\cite{cuprates1,cuprates0} particular material architectures made by heterostructures at atomic limit
which could give high-$T_c$ superconductors have been disclosed.
Panel (A) in Fig.~\ref{fig.1} shows one of these heterostructures at atomic limit: a superlattice of stripes. 
In this case the material is made of nanoscale modules,
metallic stripes of atomic thickness, as for example metallic graphene or phosphorene nanoribbons 
assembled in a superlattice of stripes separated by potential barriers from 
neighbor stripes in the same plane or in the neighbor plane.
The size of the nanoscale units or modules and the superlattice period $\lambda_p$
 is required to be in the nanoscale. In fact the Fano resonance to get the highest $T_c$ need to tune the wavelength 
 $\lambda_{Fn}$ of electrons at the Fermi level in the n-th band in the range of $\lambda_p$.
The crystalline structure of undoped p-Terphenyl~\cite{rice,lech} 
with a monoclinic space group P21/a is shown in Panel (B) of Fig.~\ref{fig.1} where gray dots are carbon atoms and red dots are hydrogen atoms, can be described as a packing of parallel p-Terphenyl nanoribbons or stripes of about 1.4 nm width running in the 
$\bf{b}$-axis direction indicated by the black arrows. The figure shows the similarity of the arrangements of the p-Terphenyl nanoscale stripes with the schematic drawing of the superlattice of stripes 
shown in panel (A) of Fig.~\ref{fig.1} from ref~\cite{cuprates0}. 
The projections in the $\bf{ac}$, $\bf{bc}$ and $\bf{ba}$ crystal planes are shown in Fig.~\ref{fig.2}. 
Two parallel stripes are clearly seen in the $\bf{bc}$ projection. 
The $\bf{ba}$ projection in Fig.~\ref{fig.2} shows a variable torsion angle (up to 14 degrees) of the middle ring relative with the two lateral ones. While the $H - H$ bond is 0.32 nm between the lateral rings of the neighbor molecules very short $H - H$ intermolecular bonds (dashed lines connecting neighbor molecules in the stripe in the $\bf{b}$ direction) are established for the central benzene ring, as short as 0.22 nm, like the shortest values observed in the high pressure phases of benzene \cite{wen,katru}. The stripes are not connected with the neighbor stripes also in the a-axis direction indicating the quasi one dimensional electronic structure of the stripes.

\begin{figure}
\onefigure[scale=1]{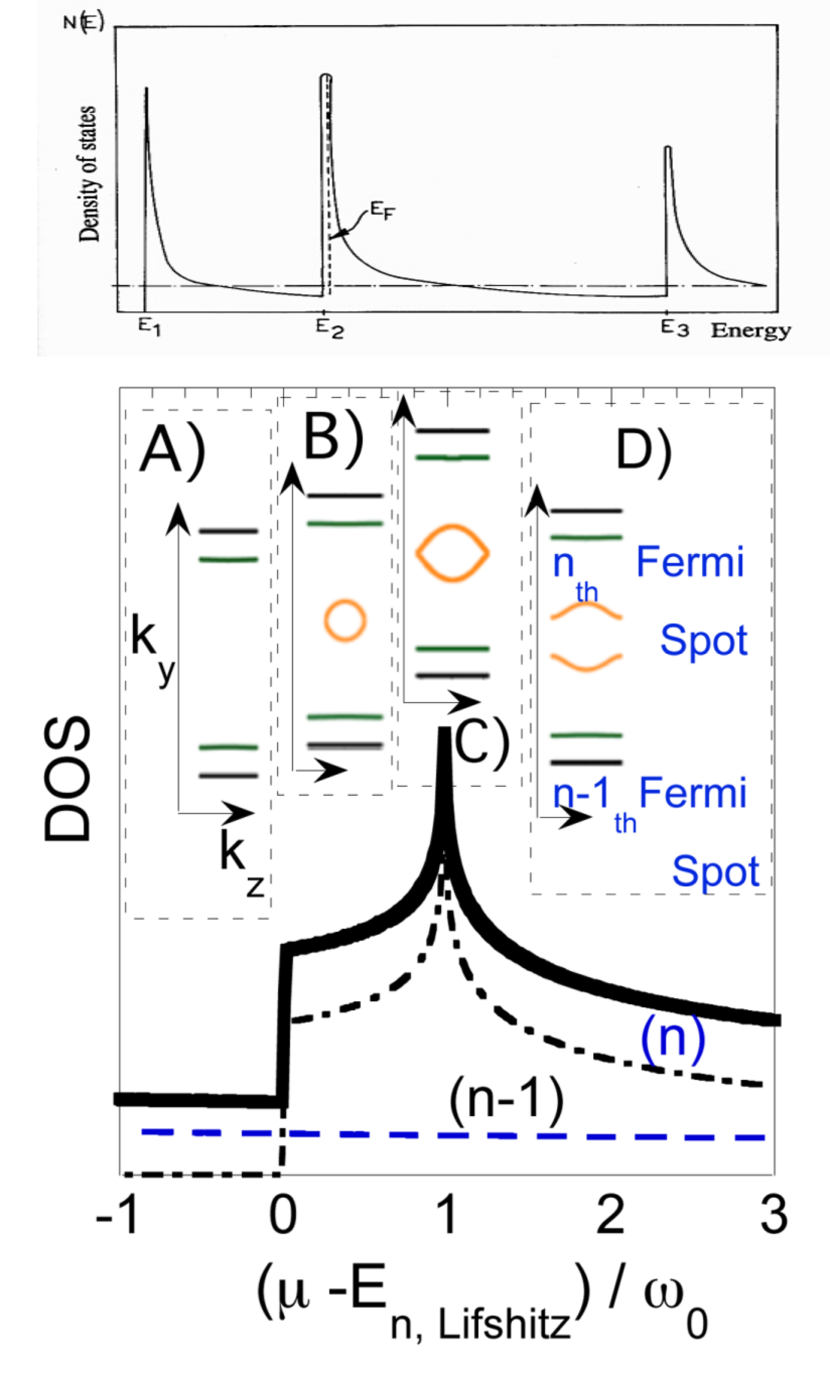}
\caption{The density of states (DOS) for the superlattice of stripes which simulate the DOS at the Van Hove singularity associated with a 1st Lifshitz transition for the appearing of a new circular Fermi surface and a 2nd 2D-1D topological Lifshitz transition. Panels (A), (B), (C) and (D) show the evolution of the Fermi surfaces arcs and pockets as the chemical potential is increased through the Lifshitz transition, in the energy range of $4\omega_0$. When the chemical potential moves from below to above the bottom of the n-th band, i.e., at the first Lifshitz transition for the appearing of a new Fermi surface, the pocket with 2D character  appears in panel (B) Continuing to increase the chemical potential, at the second Lifshitz transition, at the Van Hove singularity, panel (C) , the Fermi surface pocket opens a neck and the new Fermi surface acquires a 1D behavior forming Fermi arcs, panel (D).}
\label{fig.4}
\end{figure}

The electronic structure of p-Terphenyl~\cite{pusc,koller,fa} confirms the presence of one dimensional electronic states
in the conduction band. A portion of the electronic structure of the conduction band~\cite{pusc} is shown in the upper part of Fig.~\ref{fig.3}. The band structure, upper panel of Fig.~\ref{fig.3}, displays in a clear way a strong
anisotropic character in the wave-vector space.
Indeed, in the irreducible representation of the band structure reported in the upper part 
of Fig.~\ref{fig.3}~\cite{pusc,koller,fa} shows
two directions in the Brillouin zone for which the band dispersions
are very narrow. A narrow bandwidth of the order of hundred meV,
indicates that large potential barriers have
to be penetrated by the electrons to tunnel from one p-Terphenyl stripe to its neighbor stripe.
Interestingly and relevantly for this work, all the bands of the band-bundle due to the different
overlapping orbitals of the p-Terphenyl, have this property and are very narrow along these
two independent directions, indicating that the potential barrier formed by the crystal structure
and by the characteristics of the electronic bonds is felt by all the conduction and valence electrons. In the b-axis direction in the real space, orthogonal to the upper surface of the Brillouin zone the band
dispersion is much broader, of the order of 700 meV, pointing toward a quasi-free
conduction of the electrons with a moderate effective mass along the b direction.
The energy scale of the pairing,145 meV,
is taken from the relevant phonons detected at 1171 cm-1,
due to a mode of the C-C bond, and/or the mode at 1222 cm-1 of the C-H mode, detected by Raman spectroscopy in metallic $K_3$ p-terphenyl~\cite{a}. 
We have simulated the band dispersion and the DOS with a periodic potential barrier with 1.4 nm 
periodicity which reproduces the narrow band dispersion in the transversal direction of the 
stripes shown in the lower panel of Fig.~\ref{fig.3}. We have obtained an electronic structure
 which grabs the key feature of the evolution of the electronic structure where the chemical 
 potential is tuned at a band edge in a superlattice of stripes as shown in the upper
  panel of Fig.~\ref{fig.4} showing the drawing of the patent for material design of heterostructures at atomic limit~\cite{cuprates0}.

Once doped by K, the chemical potential will be raised in the conduction band and depending of the doping
(and/or misfit strain, pressure, orientational disorder, magnetic field ...) at the n-th band edge giving rise to a complex network of Fermi surfaces, with electron and hole-like small pockets of Fermi surfaces and Fermi arcs.
The key point of this work is to predict the superconducting properties of the K-doped
p-Terphenyl with numerical calculations through a simplified model of its nanoscale structure getting the key electronic structure near a band edge. The bands and wave-functions are created by using a proper superlattice of
stripes, which allow the solution of the Bogoliubov gap equations \cite{ann,bab} 
without standard BCS approximations.
giving high-$T_c$ superconducting state with multigaps
and multi-condensates at different BCS/BEC pairing regimes.
The pairing is thought to be driven by an attractive interaction within each Fermi surface, 
but the non-diagonal interaction between condensates can be either repulsive or attractive. 
The multicomponent character of the pairing and the geometry of the system will determine shape resonances in the
superconducting gaps and $T_c$, with peculiar features predicted 
for the isotope effect and the gap to $T_c$ ratios, which
can be tested in future experiments, together with the expected 
anisotropic transport of electrons in doped (or out of equilibrium) single crystals.
The presence of high temperature superconducting domes where the chemical potential 
crosses the Lifshitz transitions has now well confirmed by experiments in iron based superconductors.
Here we present the possible scenario of high temperature superconductivity in p-Terphenyl where the chemical potential
is driven by potassium doping at a Lifshitz transition. 
Our model predict that at the Lifshitz transition the Fermi surface is made of multiple components: circular Fermi surface
pockets and Fermi arcs as shown in Fig.4.
The superconducting properties have been calculated using the Bianconi-Perali-Valletta (BPV) theory which has been used to predict the mechanism driving the emergence of high $T_c$ in 
cuprates~\cite{shape1,shape12,isotope1,annette}, 
where the 1996 proposed scenario of the coexistence 
of Fermi arcs and Fermi pockets was confirmed in 2009 ~\cite{arcs-pockects}.
It has been able to predict the anomalous deviation of the isotope 
coefficient from the BCS predicted value~\cite{isotope3,isotope4}.
The BPV theory
has been applied to diborides~\cite{mgb0,mgb3,mgb7}, 
and iron based superconductors~\cite{iron1,iron2,iron3} where it has been 
confirmed by recent experiments~\cite{iron4,iron5,iron6,iron7,iron10}. 
In these experiments it has been reported compelling evidence that
 the high $T_c$ domes occur in the proximity
 of the Lifshitz transitions for the appearing of a new Fermi surface pocket.
Recently similar scenarios have been proposed for pressurized sulfur hydrides~\cite{annette2} and in superconducting nanofilms~\cite{doria}

 \begin{figure}
\onefigure[scale=1.1]{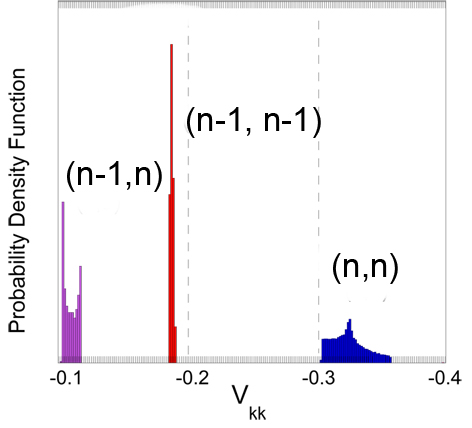}
\caption{Matrix elements of the effective pairing interaction for electrons of the (n-1)-th and n-th bands, calculated from the overlap integral of the single-particle wave-functions of the superlattice potential. Matrix elements for the intraband and exchange-pair processes are reported, demonstrating an interesting wave-vector dependence for the two-particle scatterings involving the electrons of the upper n-th band.}
\label{fig.5}
\end{figure}

\begin{figure}
\centering
\onefigure[scale=0.9]{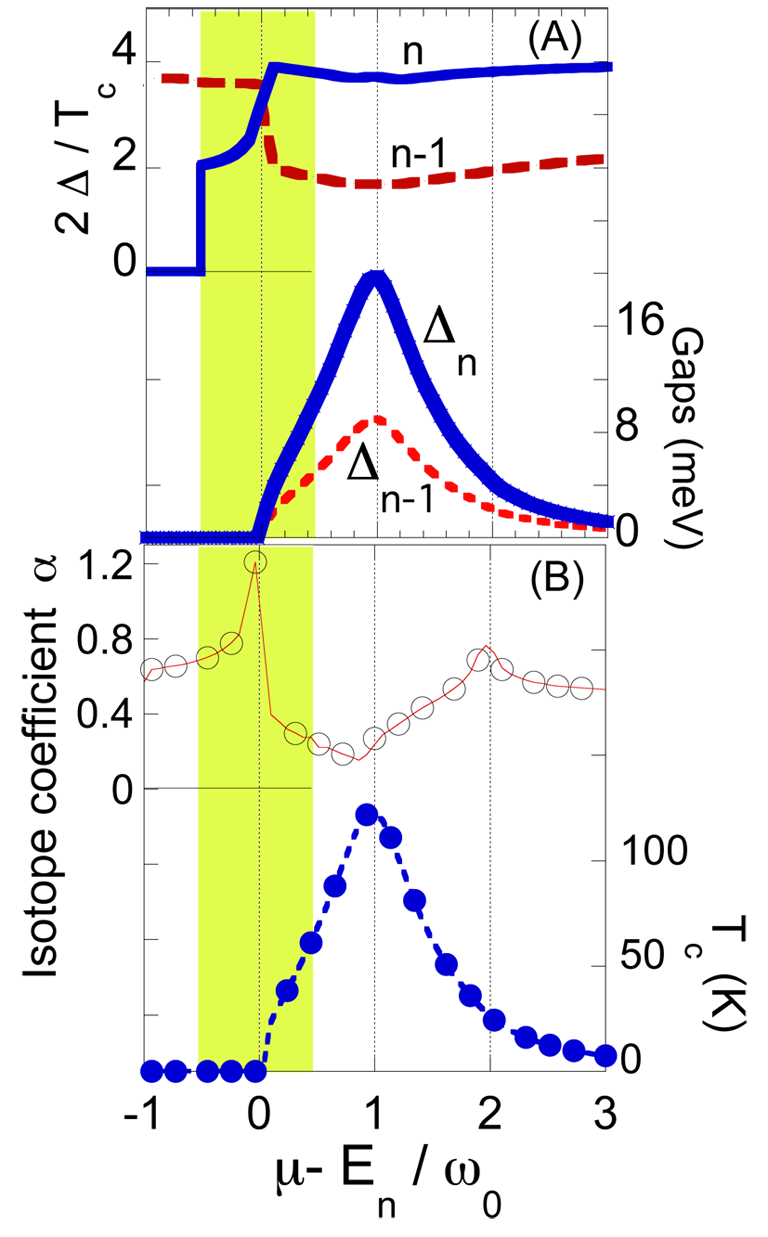}
\caption{Results of the BPV theory for the superconducting state properties as a function of the energy distance from the 2D-1D Lifshitz transition. Panel (A). Superconducting gaps at zero temperature which open in the band n and n-1, and gaps to $T_c$ ratios. Panel (B). Superconducting critical temperature and corresponding isotope coefficient. The maximum gaps and $T_c$ are located at the 2D-1D dimensional crossover, with a broad resonance induced by the large value of the phonon energy $\omega_0$=145 meV.}
\label{fig.6}
\end{figure}

Let us consider a system made of multiple bands with index $n$. The energy separation between the chemical potential 
and the bottom of the n-th band defines the Fermi energy of the n-th band. 
This formulation was proposed for systems with a band crossing the chemical potential
having a steep free electron like dispersion in the $x$ direction and a flat band-like dispersion in the $y$ direction.
The wavefunctions of electrons at the Fermi level are calculated using a lattice with quasi-one dimensional lattice 
potential modulation where the chemical potential is tuned near a Lifshitz transition, like in magnesium
diborides, A15, cuprates, iron based superconductors and we propose here for doped p-terphenyl.

The superconducting critical temperature $T_c$ in the BPV theory is obtained by numerical solution of the gaps equation \cite{ann,bab} considering the simplest case of a two dimensional system \cite{isotope1} of stripes,
but the extension to three dimensional system \cite{mgb7} is straightforward. 
The $T_c$ is determined by solving the following
selfconsistent system of equations, 

\small{\begin{equation} 
\Delta_{n,k_y}=
-\frac{1}{2N}\sum_{n',\mathbf{k'}}V_{\mathbf{k},\mathbf{k}'}^{n,n'}
\frac{\tanh(\frac{E_{n,k'_y}+\epsilon_{k_x}-\mu}{2T_c})}{E_{n,k'_y}+\epsilon_{k_x}-\mu}\Delta_{n',k'_y},
\end{equation}}
where the $E_{n,k_y}+\epsilon_{k_x}$ is the energy dispersion and $\mu$
the chemical potential. 

We consider a superconductor with multiple gaps $\Delta_{n,k_y}$ in multiple bands
$n$ with flat band-like dispersion in the y direction and steep free-electron-like 
dispersion in the x direction for a simple model of a two dimensional metal with a one-dimensional 
superlattice modulation in the y-direction. The self consistent equation for the gaps at $T=0$ where each gap depends
on the other gaps is given by

\begin{equation}
\Delta_{n,k_{y}}=-\frac{1}{2N}\sum_{n',k'_y,k'_x}
\frac{V_{\mathbf{k},\mathbf{k}'}^{n,n'}\Delta_{n',k'_y}}{
\sqrt{(E_{n',k'_y}+\epsilon_{k'_x}-\mu)^2+\Delta_{n',k'_y}^2}},
\end{equation}

where N is the total number of wave-vectors in the discrete summation
and $V^{n,n'}_{\mathbf{k},\mathbf{k}'}$ is the effective pairing interaction 
taken in the separable and energy cutoff approximation,

\small{\begin{equation}
V_{\mathbf{k},\mathbf{k}'}^{n,n'}=
\widetilde{V}_{\mathbf{k},\mathbf{k}'}^{n,n'} \times\theta(\omega_0-|E_{n,k_y}+\epsilon_{k_x}-\mu|)\theta(\omega_0-|E_{n',k'_y}+\epsilon_{k'_x}-\mu|).
\end{equation}}

\begin{figure}
\centering
\onefigure [scale=0.8] {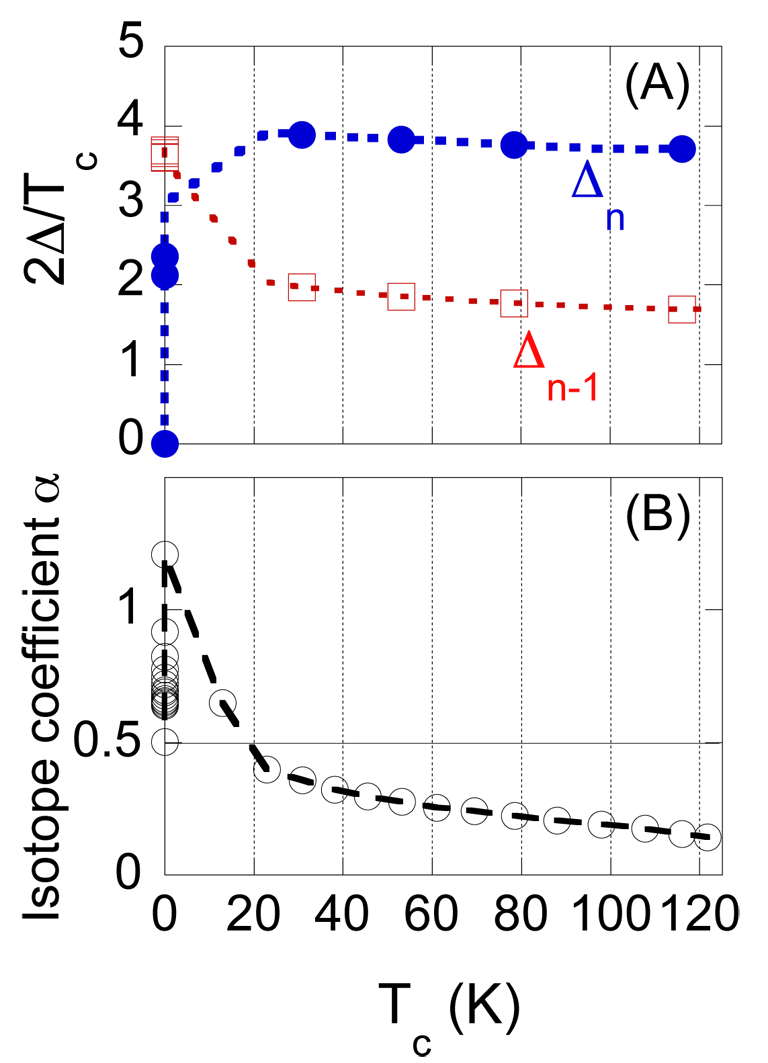}
\caption{Superconducting state properties reported as a function of $T_c$ predicted by the BPV theory. Panel (A). Zero temperature superconducting gaps to $T_c$ ratios. Panel (B). Isotope coefficient for the isotope effect on the critical temperature. Different $K$ doping levels, not well under control in the experiments, will correspond to different critical temperatures, which can be used to find the associated superconducting properties in the figure.}
\label{fig.7}
\end{figure}

Here we take account of the interference effects between the 
single-particle wave functions of the pairing electrons in the different
bands, where $n$ and $n'$ are the band indexes, $k_y(k_y')$ is the 
superlattice wave-vector and $k_x(k_x')$ is the component of the
wave-vector in the free-electron-like direction of the initial (final) state in the pairing process. 

\begin{equation} 
\begin{split}
\widetilde{V}_{\mathbf{k},\mathbf{k}'}^{n,n'} = & - \frac{\lambda_{n,n'}}{N_0}S\times \\
& \times\int_{S}
\psi_{n',-k'_y}(y)\psi_{n,-k_y}(y)
\psi_{n,k_y}(y)\psi_{n',k'_y}(y)dxdy,
\end{split}
\end{equation}

$N_0$ is the DOS at the Fermi energy $E_F$
without the lattice modulation, $\lambda_{n,n'}$ is 
the dimensionless coupling parameter, $S=L_xL_y$ is the
surface of the plane and $\psi_{n,k_y}(y)$ are the eigenfunctions in the 1D
superlattice. 
We have used weak coupling intraband coupling constants 0.14  for the (n-1)-th bands, 0.2 for the appearing n-th band and a small 0.1 interband  attractive or repulsive exchange  interband pairing constant and the distribution of calculated intraband pairing terms  (n,n) (n-1,n-1) together with the  (n,n-1) interband terms is shown in Fig.~\ref{fig.5}.

The system of equations for the gaps need to be 
solved iteratively. The anisotropic gaps depend on 
the band index and on the
superlattice wave-vector $k_y$. According with Leggett\cite{leg} 
the ground-state BCS wave function corresponds to an ensemble
of overlapping Cooper pairs at weak coupling (BCS regime) and evolves to 
molecular (non-overlapping) local pairs with bosonic
character in the BEC regime. This approach remains valid also if a particular band is in the BCS-BEC crossover and beyond Migdal approximation
because all other bands are in the BCS regime and in the Migdal approximation.
In this crossover regime the chemical potential $\mu$ 
results strongly renormalized with respect to the
Fermi energy $E_Fn$ of the non interacting system.
In the case of a Lifshitz transition, as described in this paper,
nearly all electrons in the new appearing Fermi surface condense forming a
condensate having a BCS-BEC crossover character. 

Therefore at any chosen value of the
charge density $\rho$ for a number of the occupied bands $N_b$,
the chemical potential in the superconducting phase
has to be renormalized by the gap opening solving the following equation:

\begin{equation}
\rho =\frac{1}{L_xL_y}\sum_{n}^{N_b}\sum_{k_x,k_y}\left[1-
\frac{E_{n,k_y}+\epsilon_{k_x}-\mu} {\sqrt{(E_{n,k_y}+\epsilon_{k_x}-\mu)^2+\Delta_{n,k_y}^2}}\right].
\end{equation}

Three distinct regimes of multi-gap superconductivity are obtained as a function of the chemical potential tuned around the 2D-1D Lifshitz
transition, as reported in Fig.~\ref{fig.6}. At the n-th band bottom, when the new Fermi surface pocket starts to appear, there is
a coexistence of a BCS-like condensate of the pairs of the (n-1)-th band together with a BEC-like condensate of the pairs of the n-th band, which
because of the very low density condense completely in a bosonic liquid. In this regime the critical temperature is extremely low and small
variations of the parameters can lead to large variations in the gaps and in the $T_c$, determining a large peaked value of the isotope coefficient.
The peak in the isotope coefficient reported in panel (B) is a finger print of our BPV theory and it signals the strong interplay between
the Lifshitz transition at the band bottom and the onset of shape resonant superconductivity. Increasing the chemical potential trough the
second Lifshitz transition, the resonant regime of maximum $T_c$ and gaps is obtained. This resonant regime is characterized by an interesting
coexistence of BCS-like pair condensate of the (n-1)-th band and BCS-BEC crossover-like pair condensate of the n-th band. In the resonance regime
the two gaps differ by a sizable 2.5 factor, the isotope effect gets its smallest values and $T_c$ can reach the high value of 123 K with coupling
strengths in the different channels not exceeding 0.3, values still typical of metals, as Nb. The large energy scale of the phonon (145 meV)
determines not only a large prefactor for the critical temperature, but also induces a large width of the resonant regimes, making relatively
simple to find the right doping levels giving the highest critical temperatures. Finally, for larger chemical potentials, a third regime
of conventional two-band superconductivity is reached, with coexistence of two-particles condensates having both BCS-like character, confirmed
also by values of the isotope coefficient approaching 0.5 and small values of the gaps typical of weakly-coupled superconductors.
The predicted zero temperature superconducting gaps to $T_c$ ratios, panel (A), and the Isotope coefficient as function on the critical temperature, panel (B) are reported in Fig.~\ref{fig.7} so that the present theoretical work can be confirmed by direct experiments.
This theory does not include superconducting fluctuations. 
However it should be noticed that in a multigap / multiband system
another fundamental phenomenon helps in stabilizing high
temperature superconductivity:
the screening of the superconducting fluctuations. 
In fact the multi-band BCS-BEC crossover in a two-band 
superconductor (one condensate in the BCS regime, the other in the BCS-BEC
regime) can determine the optimal condition to allow the screening of the
detrimental superconducting fluctuations due to the large stiffness
of the BCS-like condensate in the deep band. 
 Finally an arrested or frustrated phase separation is expected to occur at a topological Lifshitz transition~\cite{kugel1,kugel2} as it was observed in cuprates, and diborides~\cite{campi1,saini1}, diborides~\cite{campi2}. 

\acknowledgments
This work has been supported by superstripes-onlus. Discussions with Gianni Profeta, Augusto Marcelli, Nicola Poccia, Andrea Ienco, Corrado Di Nicola, Fabio Marchetti, Claudio Pettinari are acknowledged.

\end{document}